# Dynamic magnetic response of infinite arrays of ferromagnetic particles

Kirill Rivkin,[1] W. Saslow,[1] V. Chandrasehkhar,[2] L. E. De Long[3] and J. B. Ketterson[2]

[1] Department of Physics, Texas A &M University, College Station, TX 77843

[2] Department of Physics and Astronomy, Northwestern University, Evanston IL, 60208

[3] Department of Physics University of Kentucky, Lexington, KY 40506-0055

**Abstract**

**Recently developed techniques to find the eigenmodes of a ferromagnetic particle of arbitrary shape, as well as the absorption in the presence of an inhomogeneous radio-frequency field, are extended to treat infinite lattices of such particles. The method is applied to analyze the results of recent FMR experiments, and yields substantially good agreement between theory and experiment.**

## 1. Introduction

Studies of the resonant modes of magnetic systems have been conducted for over 70 years, dating from the work of Landau and Lifshitz. Currently most experiments involve either Ferromagnetic Resonance (FMR) (in which a magnetic sample is subjected to a RF field with a fixed frequency while monitoring the energy absorption as the magnitude and direction of the external DC magnetic field are varied), or Brillouin Light Scattering (in which the frequency shift arising from inelastic scattering from thermally excited spin-wave modes is monitored for a sample illuminated by laser light). For ellipsoidal bodies and limiting shapes there of (e.g., cylinders and discs), analytical solutions for the resonant frequencies can be obtained, allowing comparisons between theory and experiment. However, such solutions are unattainable for the majority of sample shapes, with very few exceptions such as the vortex gyroscopic mode in discs[1].

The mode structure of magnetic nanoparticles presents another complication in that experiments are typically done on patterned arrays containing tens of thousands of such particles single particles, rather than single particles, in order to obtain a stronger signal that can be easily measured (we note in passing that arrays are also relevant to hard-disc information storage systems that typically involve of order $10^{12}$ nanoparticles[2]). Until now, numerical simulations of such arrays were generally deemed to be impossible; however, the approach introduced below allows some array properties to be calculated.

Recently a method was developed to calculate the modes of a ferromagnetic particle having an arbitrary shape.[3] It was based on the assumption that a body can be represented by an array of macro-spins, each consisting of many true spins, which fill that body on a specified spatial grid (an approximation sometimes called the "discrete dipole approximation"). The spin dynamics is then assumed to follow the Landau-Lifshitz (LL) equation, which we write as[4]

$$\frac{d\mathbf{m}_i}{dt} = -\gamma \mathbf{m}_i \times \mathbf{h}_i^{total} - \frac{\beta\gamma}{M_s} \mathbf{m}_i \times \left(\mathbf{m}_i \times \mathbf{h}_i^{total}\right); \qquad (1.1)$$

where $\gamma$, $M_s$, and $\beta$ are the gyromagnetic ratio, saturation magnetization, and (dimensionless) damping parameter, respectively, and $\mathbf{h}_i^{total}$ is the total effective field experienced by the i$^{th}$ macro-spin in the array. Each spin feels the effect of the others through the long-range dipole-dipole field $\mathbf{h}_i^{dipole}$, and the short-range exchange field $\mathbf{h}_i^{exchange}$; in addition they interact with the external magnetic field $\mathbf{H}_0$, and, if present, an anisotropy field $\mathbf{h}_i^{anisotropy}$ that we will neglect in what follows. Hence, we have $\mathbf{h}_i^{total} = \mathbf{h}_i^{dipole} + \mathbf{h}_i^{exchange} + \mathbf{H}_0$. We linearize Eq. (1) by writing $\mathbf{m}_i = \mathbf{m}_i^{(0)} + \mathbf{m}_i^{(1)}$ and $\mathbf{h}_i = \mathbf{h}_i^{(0)} + \mathbf{h}_i^{(1)}$. Combining these expressions and retaining only first-order terms yields

$$-\frac{d\mathbf{m}_i^{(1)}}{dt} = \gamma \left[\mathbf{m}_i^{(0)} \times \mathbf{h}_i^{(1)} + \mathbf{m}_i^{(1)} \times \mathbf{h}_i^{(0)}\right] + \frac{\gamma\beta}{M_s} \mathbf{m}_i^{(0)} \times \left[\mathbf{m}_i^{(0)} \times \mathbf{h}_i^{(1)} + \mathbf{m}_i^{(1)} \times \mathbf{h}_i^{(0)}\right]. \qquad (1.2)$$

We next assume solutions of the form $\mathbf{m}_i^{(1)}(t) = \mathbf{m}_i^{(1)} e^{-i\omega_n t}$ (solutions are always expressed as complex-conjugate pairs and are hence real). The resulting coupled equations have the structure of a *vector eigenvalue problem*. The *eigenvalues*, $\omega_n$, give the frequencies of the (generally mixed) dipolar/spin-wave modes; the number of such modes is equal to the total number of dipoles N, used to discretize the nanoparticle. As a byproduct of the diagonalization process, one also obtains the

*eigenvectors*, $\mathbf{V}_n(i)$, of the associated matrix, which give the vector amplitude and phase of each spin in the nanoparticle for a given mode, n. By setting the exchange constant J = 0 we can eliminate the spin wave modes, leaving only the Walker modes.[5]

The eigenvectors can be used to solve the inhomogeneous LL equation one has in the presence of an external drive field $\mathbf{h}_i^{rf}(t) = \mathbf{h}_i^{rf}(0)e^{-i\omega t}$ acting on the i$^{th}$ spin; the only constraint on the position dependence of the drive field is that it satisfy Maxwell's equations in the quasi-static limit (the same being true for the static external field)[6,7]. This, in turn, allows a direct evaluation of the non-local (i-, j-dependent) magnetic susceptibility, $\chi_{ij}(\omega)$, and the linear response of the system (a detailed derivation can be found in our previous work[8]). One then has

$$\mathbf{m}_i(t) = \sum_j \chi_{ij}(\omega) \cdot \mathbf{h}_j^{rf}(t), \qquad (1.3a)$$

or in vector-component form,

$$m_{i\alpha}^{(1)} = iV_{i\alpha}^{(k)} \frac{V_{Ll\beta}^{(k)*} \gamma \varepsilon_{\beta\sigma\chi} m_{l\sigma}^{(0)} h_{l\chi}^{(rf)}}{\omega^{(k)} - \omega} e^{-i\omega t}. \qquad (1.3b)$$

Here the summation convention is assumed, where Roman letters correspond to individual dipoles, Greek letters correspond to spin and field components, $V_{Ll\beta}^{(k)*}$ is a "left eigenvector" of Eq.(1.3) (required since the right eigenvectors alone do not form a complete orthonormal basis of solutions for the matrix in Eq. (1.2), which is non-Hermitian) and $\varepsilon_{\beta\sigma\chi}$ is a Levi-Civita symbol. The imaginary component of $\chi_{ij}(\omega)$ allows one to calculate the resonant absorption through the expression

$$\dot{E} = 2 \lim_{t \to \infty} \frac{1}{t} \int \sum_i \frac{dm_i}{dt} \cdot H_i dt = \omega \vartheta \gamma \, \mathrm{Re}\left( -\frac{V_{Ll\beta}^{(k)*} \varepsilon_{\beta\sigma\chi} m_{l\sigma}^{(0)} h_{l\chi}^{(rf)}}{\omega - \omega'^{(k)} + i\beta\omega''^{(k)}} V_{i\alpha}^{(k)} h_{i\alpha}^{(rf)*} \right). \qquad (1.4)$$

where $\vartheta$ is a volume of a single discretization cell.

## 2. Modes in periodic lattices of nanoparticles

Although the eigenvalue method provides a new and powerful opportunity for the numerical analysis of spin waves, it has not yet addressed the problem of *patterned arrays* of magnetic nanoparticles

Most treatments to date neglect the interaction between individual nanoparticles, because it is computationally demanding to directly simulate even small (10 × 10) arrays of magnetic nanoparticles, whereas typical experiments can involve more than 10,000 nanoparticles. Numerous attempts have been made to lift this restriction. Analytical formulas have been constructed for fields produced in an array of uniformly magnetized square nanoparticles;[9] however, such formulas are not available for nanoparticles of other shapes or for a non-uniform magnetization.

Here we propose a different method for dealing with arrays of arbitrarily shaped magnetic nanoparticles, based on two assumptions. First, we assume that for sufficiently high external magnetic fields (typically 300 Oe and above for a soft material like permalloy) all of the nanoparticles have essentially the same internal distribution of the magnetization in equilibrium, although this distribution may be nonuniform within a given particle. Second, we assume that sufficiently large arrays may be regarded as infinite, so the eigenvectors associated with the modes have the Bloch-Floquet[10] form represented as $\mathbf{V}_{\mathbf{k}n}(i)e^{i\mathbf{k}\cdot\mathbf{R}}$. Related approach was described by Giovannini et al[11]. Here $\mathbf{V}_{\mathbf{k}n}(i)$, which is strictly periodic, expresses the distribution of phases and amplitudes inside the nanoparticles, i is an index numbering all spins in a given unit cell, $\mathbf{k}$ is a continuous Bloch wave vector, and n is a band index numbering all the modes of an individual nanoparticle; $\mathbf{R} = n_a\mathbf{a} + n_b\mathbf{b}$ is the set of all two-dimensional, real-space lattice vectors where $n_a$ and $n_b$ are integers and $\mathbf{a}$ and $\mathbf{b}$ are the primitive translation vectors of the two-dimensional Bravais lattice (see Fig. 1). A mode frequency, $\omega_{\mathbf{k}n}$ is associated with each Bloch state (a similar assumption is made in studying the phonons in an infinite system with many atoms per unit cell). For our array, the fields entering Eq. (1.2) can be written as

$$h_{i\alpha}^{(0,1)} = \sum_{\beta,\mathbf{r}_j} \sum_{n_b,n_a} A_{\alpha\beta}(\mathbf{r}_i - \mathbf{r}_j - n_a\mathbf{a} - n_b\mathbf{b}) m_\beta^{(0,1)}\left(\mathbf{r}_j, n_a, n_b\right) \qquad (2.1)$$

where $h_{i\alpha}^{(0)}$ and $h_{i\alpha}^{(1)}$ are the static and dynamic fields respectively, Roman letters refer to individual dipoles in the same cell, and Greek letters to different coordinate projections; $A_{\alpha\beta}$ is a **demagnetization**

*tensor* that gives the field at point $\mathbf{r}_i$ due to a magnetic dipole at point $\mathbf{r}_j$ in the same cell and, due to the sum over $n_a$ and $n_b$, all other unit cells (in one and two dimensions these sums are convergent). Since the number of nanoparticles is typically large, one can impose periodic boundary conditions.

The *static* magnetization is assumed to be strictly periodic, i.e., $m_\beta^{(0)}(\mathbf{r}_j, n_a, n_b) = m_\beta^{(0)}(\mathbf{r}_j, 0, 0)$ for arbitrary $n_a$ and $n_b$; on the other hand for the *dynamic* fields $m_\beta^{(1)}(\mathbf{r}_j, N_a, N_b) = V_{\mathbf{kn}}(\mathbf{r}_j) e^{i\mathbf{k}\cdot(N_a \mathbf{a} + N_b \mathbf{b})}$ where $N_a$ and $N_b$ are the number of unit cells in the **a** and **b** directions over which the lattice is assumed to repeat itself. The allowed values of **k** along the **a**- and **b**-directions are then given by $k_a = 2\pi m_a / N_a$ and $k_b = 2\pi m_b / N_b$ where $m_a$ and $m_b$ are also integers. Eq. (2.1) can then be simplified by introducing an effective **k**-dependent demagnetization tensor $A_{\alpha\beta}^{\mathbf{k}}(\mathbf{r}_i - \mathbf{r}_j)$ which is written

$$A_{\alpha\beta}^{\mathbf{k}}(\mathbf{r}_i - \mathbf{r}_j) = \sum_{n_a, n_b} A_{\alpha\beta}(\mathbf{r}_i - \mathbf{r}_j - n_a \mathbf{a} - n_b \mathbf{b}) \, e^{i\mathbf{k}\cdot(n_a \mathbf{a} + n_b \mathbf{b})} \qquad (2.2)$$

where $\mathbf{k} = 0$ in Eq. (2.2) corresponds to the static demagnetization tensor. Then the eigenvalue problem given by Eq. (1.2) takes on the *same form as for a single nanoparticle*:

$$i\omega V_{\mathbf{kn}}(\mathbf{r}_i) = \gamma \left[ \mathbf{m}_i^{(0)} \times \left( \sum_{\mathbf{r}_j} \mathbf{A}^{\mathbf{k}}(\mathbf{r}_i - \mathbf{r}_j) V_{\mathbf{kn}}(\mathbf{r}_j) \right) + V_{\mathbf{kn}}(\mathbf{r}_i) \times \mathbf{h}_i^{(0)} \right] + \text{damping} \qquad (2.3)$$

with two important differences: the static and dynamic magnetic fields are now due to *all* of the nanoparticles in the array and, in addition, depend upon the value of the wavevector **k**; i.e., we must do a separate diagonalization for each value of the wave vector. In spite of this, the formalism represents a great computational simplification over that required for finite arrays where Bloch's theorem does not apply.

## 3. Applications to periodic disc arrays

We now use the above formalism to analyze the results of an experiment[12] performed on a 400 × 400 micrometer array of Permalloy nanoparticles, 500nm diameter, 85nm thick, and 600nm between centers (see Figure 2). The external DC field is applied *parallel* to the array axis and also at an angle of *45°*. A uniform RF field with a frequency 9.37 GHz was applied perpendicular to the sample plane and the absorption as a function of the applied field strength measured.

It is expected that the inter-dot interaction will result in differences in the 0° and 45° spectra; however, previously we were not able to explain the physical nature of such differences – a problem we address in the present work by employing our lattice-based theory. We start by estimating which values of the wavevector **k** contribute to the absorption spectrum. Numerical calculations show that it is important to include the dipole-dipole fields due to other nanoparticles when the distance between them is less than about 20 micrometers. This implies that only modes with wavevectors greater than $k = 2\pi \cdot 10^4 / 20$ cm$^{-1}$ will be effectively distinguishable from those at $\mathbf{k} = 0$. On the other hand, such rapidly oscillating waves will not contribute appreciably to the absorption due to the fact that the applied RF field is uniform. As a result, we only need to consider the modes with $\mathbf{k} = 0$ in this experiment.

Since the values for the damping and saturation magnetization of a given sample can be somewhat different from those of other samples, we developed a procedure which allows us to slightly vary such parameters without performing unnecessarily repetitious calculations. We do this by using first-order perturbation theory. As an example, we consider the linearized Landau-Lifshtiz equation and regard the damping to be a small perturbation:

$$i(\omega + \Delta\omega)\mathbf{m}_i^{(1)} = i\omega \mathbf{m}_i^{(1)} + \frac{\gamma\beta}{M_s}\mathbf{m}_i^{(0)} \times \left[\mathbf{m}_i^{(0)} \times \mathbf{h}_i^{(1)} + \mathbf{m}_i^{(1)} \times \mathbf{h}_i^{(0)}\right]$$

$$\Delta\omega = -i\beta \sum_{i,j} \mathbf{V}_{iL} \cdot \left(\frac{\gamma}{M_s}\mathbf{m}_i^{(0)} \times \left[\mathbf{m}_i^{(0)} \times \mathbf{h}_i^{(1)}(\mathbf{V}_j) + \mathbf{V}_i \times \mathbf{h}_i^{(0)}\right]\right)$$

(3.1)

In fact, the first order effect of the damping is to linearly shift the imaginary part of ω from 0 to some negative value, we can use some arbitrary but small initial value, say $\beta = 0.01$, and then simply linearly scale β to obtain different absorption spectra.

In the same way we can relate the effect of a small shift in the saturation magnetization to a shift in the resonant frequency by using

$$\Delta\omega = \frac{\Delta M_s}{M_s} \sum_{i,j} \mathbf{V}_{iL} \cdot \left( \mathbf{m}_i^{(0)} \times \mathbf{h}_i^{(1)}(\mathbf{V}_j) + \mathbf{V}_i \times \mathbf{h}_{i\,demag}^{(0)} \right) \qquad (3.2)$$

where $\mathbf{h}_{i\,demag}^{(0)}$ is a demagnetization field alone, i.e. the field produced by the magnetic nanoparticles. Such a shift is independent of the applied field, and therefore we only need to calculate the shift in the resonant frequencies once and then apply it to the whole spectrum, thereby obtaining a new absorption spectrum. This technique is of course valid only when one can, to first order, neglect the change of $\mathbf{m}_i^{(0)}$; i.e., the sample is nearly saturated due to the high external field.

Since material parameters do not depend on the direction of the applied field, we have used the 0° absorption data to fit the material parameters in the vicinity of traditional values for bulk permalloy ($M_s = 795\,\mathrm{erg/cm}$, $\beta = 0.01$), keeping in mind that in nanostructures the damping is typically greater[13] and the saturation magnetization is typically smaller than in bulk materials. We then compare the predicted spectrum with the 45° data and see if the results are reasonable. For analysis we choose a high-field region (larger than 1000 Oe) of the data, since here one can be certain that the samples are nearly uniformly magnetized. We discretize the system using individual cells of $5 \times 5 \times 85$ nm in size, with the exception of the isolated nanodot, where a relatively good convergence was obtained for a slightly coarser discretization, $7.5 \times 7.5 \times 85$ nm. The convergence was tested by calculating the absorption maxima for different cell sizes; as a criteria we required that decreasing the in-plane cell's dimensions by a factor of 1.5 shift the position of the maxima by no more than 2%. However, we were not able to do so with the sample's thickness since using finer discretization in this direction would require an extremely significant increase of the number of cells in the system, severely increasing the computation time. As a result, we are forced to neglect the presence of surface modes and the bulk modes with non-uniform distribution along the thickness. The low-frequency surface modes are typically anti-symmetric, and therefore can not couple to the uniform r.f. field, the same is not true for the bulk modes. In all cases we neglect microwave eddy currents; while it is not obvious that this can be done for such a thick sample, the comparison between the simulations and the rather than eddy currents. The fitted material parameters are $M_S = 767 \, \text{erg/cm}$ and $\beta = 0.03$.

The calculated absorption derivative along with the experimental data for the DC field applied at 0° (parallel to the array axis) and 45° are shown in figures 3 and 4 respectively. The difference between the predicted resonance field for an isolated dot (1530 Oe) and the experimental value (1265 Oe) is only partially due to the interdot field, which, when averaged across the sample, is approximately 533 Oe for the external DC field applied at 0°, and is approximately 510 Oe at 45°. In both cases this average inter-dot field is parallel to the direction of the applied DC field. Figure 4 yields rather good agreement with experiment for the uniform mode using the same parameters as used for figure 3; unfortunately the strength of the experimentally observed signal is limited by a recorder cutoff. There are two distinct differences between these graphs: first, the shift of the uniform mode frequency with respect to that of a single dot, and second, additional high-field absorption peaks are present (three in figure 4, one in the case of a single dot, and none when the DC field is applied at 0°).

Both of these effects must be related to the spatial distribution of the field produced by nanoparticles in the array. The static inter-dot field distribution (i.e. the field inside the dot due to the other dots), as represented by the spin orientation (which lies parallel to the local field), is shown in figure 5 for an external field of 1480 Oe; it changes relatively little with changes in the external field. The spatial distributions of the excited modes themselves are shown in figures 6 and 7.

When the external d.c. field is applied at 0°, the inter-dot interaction makes the fields near the perimeter more non-uniform than in the center (figure 5a)), while pushing the "uniform" mode to higher frequencies, and therefore lower fields (Fig. 6a) vs. Fig. 6b)); the basic set of modes remains virtually unchanged – a high field satellite peak that was observed at 2500 Oe in the isolated dot (Figure 7a)), is virtually the same as the one excited here, albeit at a much lower field of 1375 Oe (figure 7b)). However, when the external DC field is applied at 45°, the situation is somewhat different, as shown in figure 5b). The inter-dot field then creates three areas where the magnetization is almost uniform – one in the dot center and two in the corners, separated from each other by a non-uniform magnetization region. This not only affects the uniform mode (figure 6c)), but also leads to the creation of a number of localized "edge modes" that differ from each other by the number of nodes, and exist in the above-mentioned areas of nearly uniform magnetization (figures 7c) – 7e)).

The difference in the position of the experimental and numerical fields of the three high-field peaks (approximately 5% error) may arise from an observed, slightly non-circular shape of the dots[14], where the latter results in observed deviations from a precise four-fold symmetry in our experiments. Since the high-field peaks are edge mode resonances, they will be more strongly affected by edge imperfections compared to the uniform mode. Also one can see that in Fig. 1 experimental curve contains a broadening to the left of the main peak, which means the presence of extra absorbing modes in the system. From preliminary calculations we may conclude that this difference between our numerical analysis and the experiment is due to the fact that we neglected the presence of the modes with a variation along the thickness, which are responsible for the appearance of this "shoulder".

## 3. Conclusions.

In the present paper we have shown how the translational symmetry of periodic systems, resulting in the Bloch-Floquet form, can be applied to analyze the modes of lattices of magnetic nanodots. For modes with wavevector $\mathbf{k} = 0$, this technique requires approximately the same number of calculations as that needed to analyze the properties of a single magnetic dot, and produces a reasonably good agreement with existing experimental data. The latter allows us to analyze the spectra of magnetic nanodots, including specific changes in the absorption spectra due to the presence of significant inter-dot interactions.

The work was supported by DOE Grants DE-FG02-06ER46278 and DE-FG02-97ER45653, and by the National Science Foundation under grants ESC-02-24210 and 03-29957.

---

[1] K. Yu. Guslienko and Slavin, Journal of Applied Physics **87**, 6337 (2000).

[2] As discussed in http://www.cdrinfo.com/Sections/News/Details.aspx?NewsId=18034, a news release by Hitachi.

**Figure Captions.**

**Fig.1. Array of nanodots.**

**Fig. 2. SEM image of a periodic are of perm alloy discs.**

**Figure 3. The heavy line shows the experimental absorption derivative as a function of the external d.c. field. The simulations for isolated dot and the dot array are shown by the dashed and continuous lines, respectively; the d.c. field is applied at 0° with respect to the array axis.**

**Figure 4.** Absorption derivative (heavy line) and the simulation (light line) as a function of the external d.c. field for the dot array (the saturation at the lower extreme is an instrumental effect); the d.c. field is applied at 45° with respect to the array axis.

**Figure 5.** Local spin orientations arising from the external field and interdot fields (associated with other nanoparticles); a) external DC field is applied at 0°, and b) at 45 degrees.

**Figure 6.** Z-projections of uniform modes for a) an isolated dot (1530 Oe), b) an array with the external field at 0° (1265 Oe), and c) 45° (1170 Oe). The magnitude of excitations is color coded.

**Figure 7.** Z-projections of low-frequency resonant modes (those responsible for high-field satellites). a) - an isolated dot (2500 Oe), b) - 0° (1375 Oe) and c) - e 45° (2097, 2345 and 2630 Oe) arrays. The magnitude of excitations is color coded.

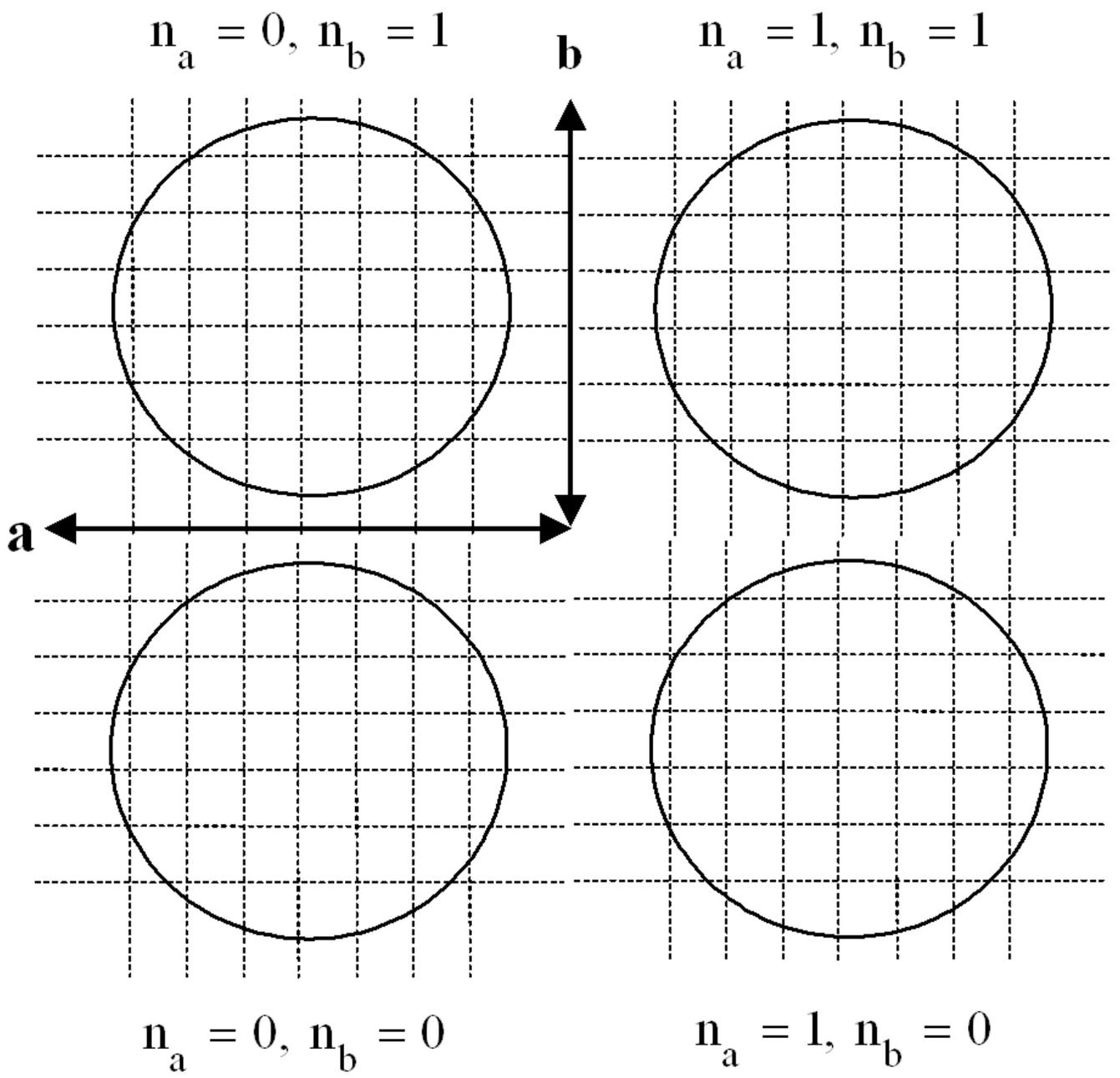

**Figure 1.**

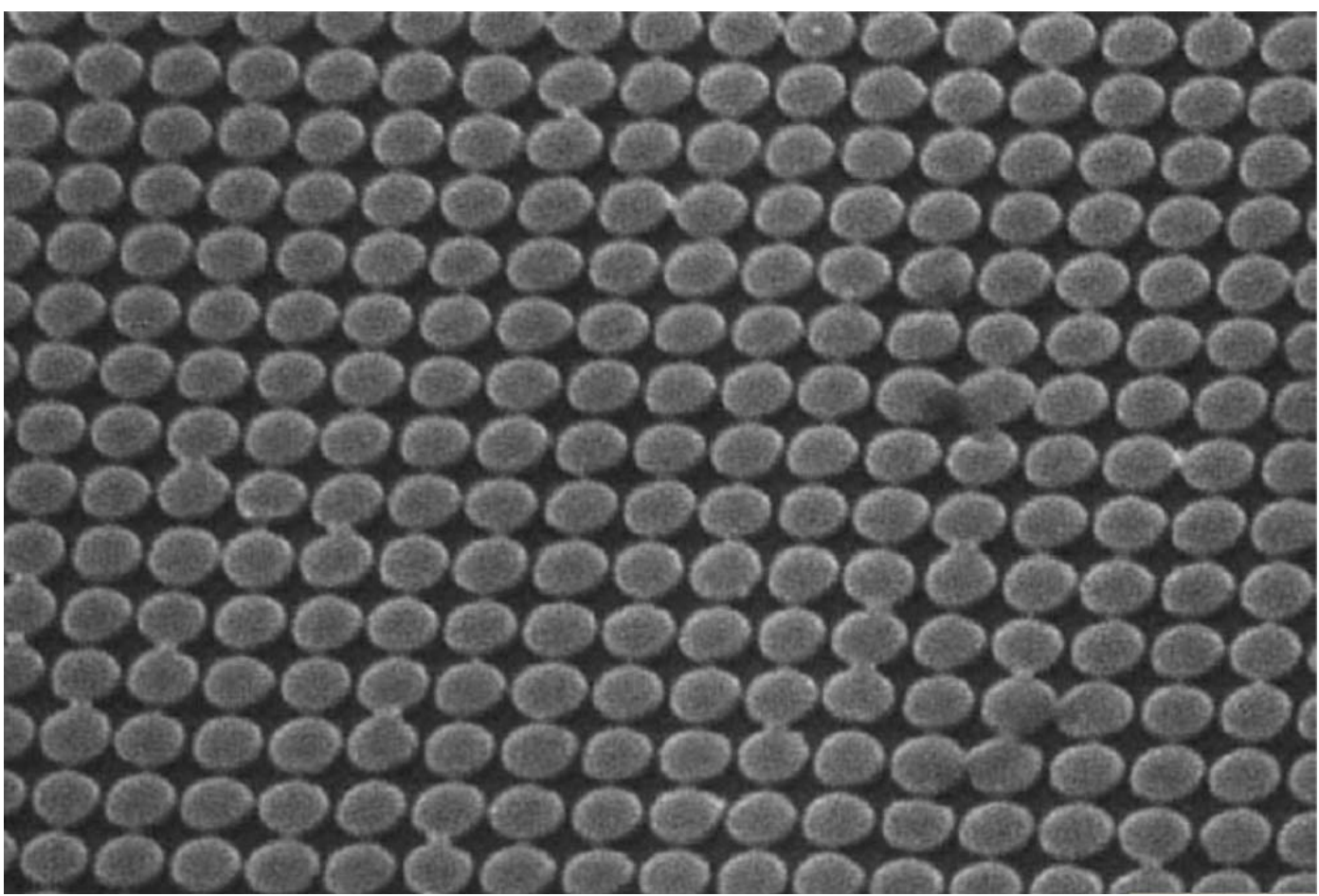

Figure 2.

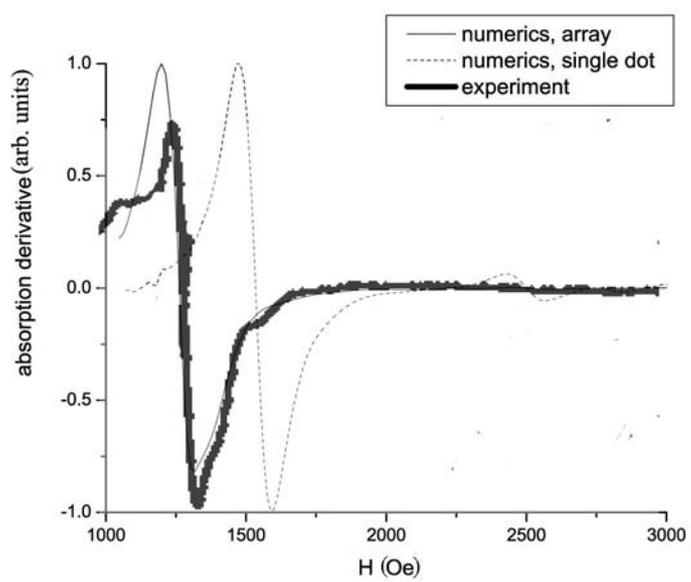

Figure 3.

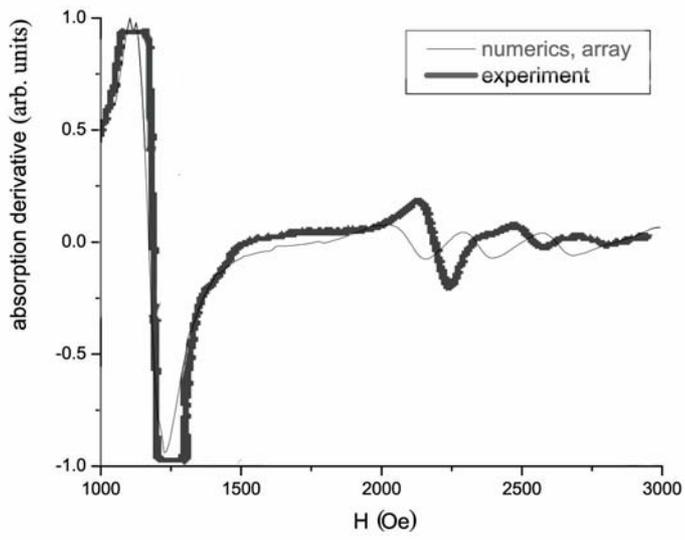

Figure 4.

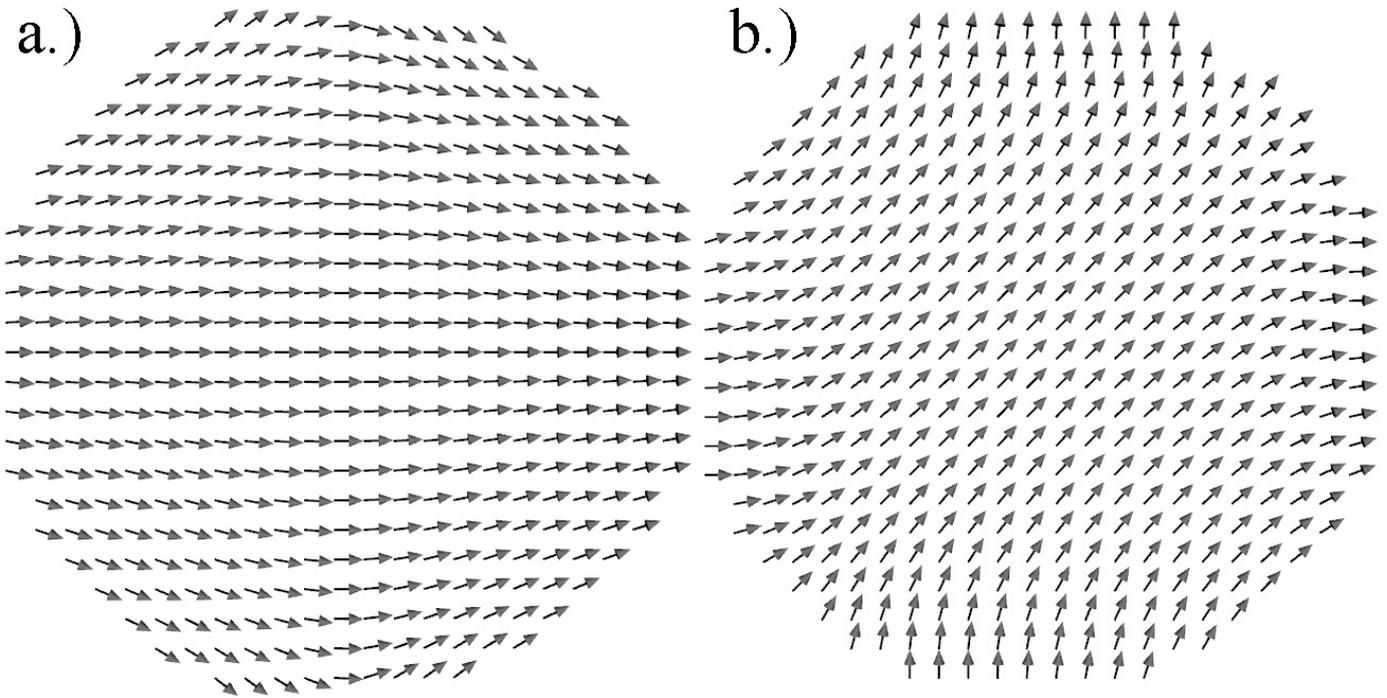

Figure 5.

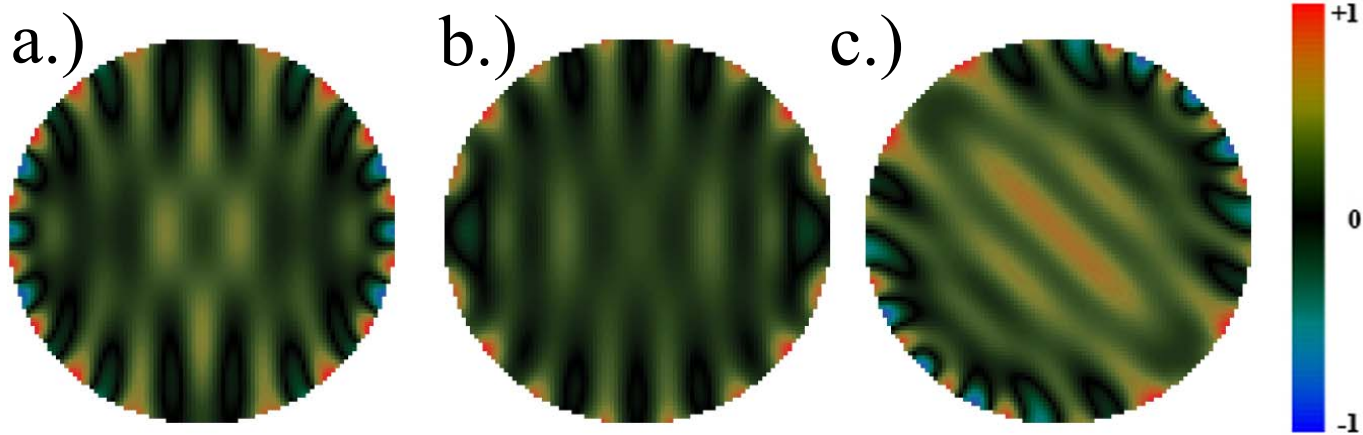

Figure 6.

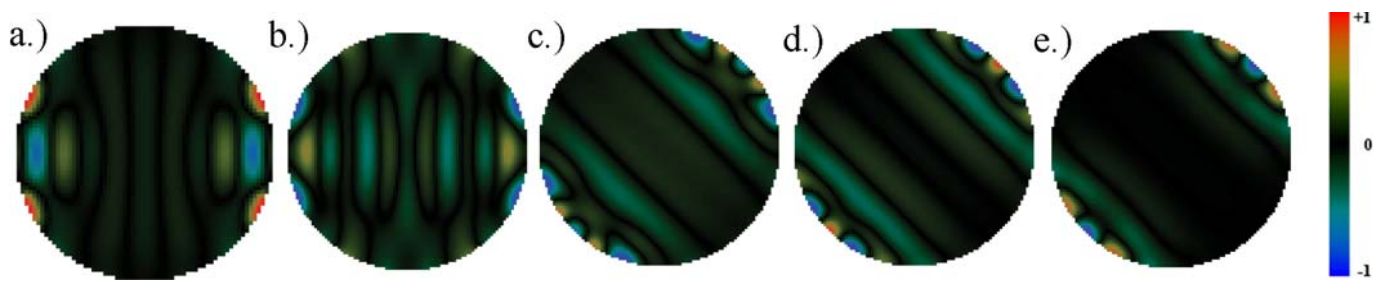

Figure 7.